\documentclass[sigconf]{cidr-2025}
\usepackage{todonotes}
\setuptodonotes{inline}

\usepackage{tabularx}
\usepackage{fancyvrb}
\usepackage{subfig}
\usepackage{xcolor}

\AtBeginDocument{%
  }

\begin{document}

\title{The Design of an LLM-powered Unstructured Analytics System}

\author{Eric Anderson, Jonathan Fritz, Austin Lee,
Bohou Li, Mark Lindblad, Henry Lindeman, Alex Meyer, Parthkumar Parmar,
Tanvi Ranade, Mehul A. Shah, Benjamin Sowell, Dan Tecuci, Vinayak Thapliyal,
Matt Welsh
\\ Aryn, Inc.}

\renewcommand{\shortauthors}{Anderson et al.}

\begin{abstract}

  LLMs demonstrate an uncanny ability to process unstructured data,
  and as such, have the potential to go beyond search and run complex,
  semantic analyses at scale. We describe the design of an {\em
    unstructured analytics} system, Aryn, and the tenets and use cases
  that motivate its design. With Aryn, users specify queries in
  natural language and the system automatically determines a {\em
    semantic} plan and executes it to compute an answer from a large
  collection of unstructured documents. At the core of Aryn
  is {\em Sycamore}, a declarative document processing engine, that provides 
  a reliable distributed abstraction called {\em DocSets}. Sycamore allows users to
  analyze, enrich, and transform complex documents at scale. Aryn includes
  {\em Luna}, a query planner that translates natural language
  queries to Sycamore scripts, and {\em DocParse}, which takes
  raw PDFs and document images, and converts them to DocSets for
  downstream processing. We show how these pieces come together to achieve better accuracy
  than RAG on analytics queries over real world reports from the
  National Transportation Safety Board (NTSB). Also, given current limitations of LLMs, we argue that an analytics system must provide explainability to be practical, and show how Aryn's user interface does this to help build trust.

\end{abstract}
\maketitle

\section{Introduction}

Large language models have inspired the imagination of industry, and companies are
starting to use LLMs for product search, customer support chatbots, code
co-pilots, and application assistants. In enterprise settings,
accuracy is paramount. To limit hallucinations, most of these
applications are backed by semantic search architectures that answer
queries based on data retrieved from a knowledge base, using
techniques such as {\em retrieval-augmented generation (RAG)}~\cite{rag-2021}. 

Still, enterprises want to go beyond RAG and run
{\em semantic} analyses that require complex reasoning across large
repositories of unstructured documents. For example, financial
services companies want to analyze research reports, earnings calls, and
presentations to understand market trends and discover investment opportunities.
Consumer goods firms want to improve their marketing strategies by
analyzing interview transcripts to understand sentiment towards brands.
In legal firms, investigators want to analyze legal case summaries to
discover precedents for rule infringement and the actions taken
across a broad set of companies and cases.

In addition to simple ``hunt and peck'' queries, for which RAG is 
tailored, these analyses often require ``sweep and harvest'' patterns.
An example is a query like ``What is the yearly revenue growth and
outlook of companies whose CEO recently changed?''
For this, one needs to sweep through large document collections, perform a mix of
natural-language semantic operations (e.g.,
filter, extract, or summarize information) and structured operations
(e.g. select, project, or aggregate), and then synthesize an
answer. Going a step
further, we see ``data integration'' patterns where users want
to combine information from multiple collections or sources. For example,
``list the fastest growing companies in the BNPL market and their competitors,'' where
the competitive information may involve a lookup in a database
in addition to a sweep-and-harvest phase to gather the top companies.
We also expect complex compositions of these patterns to become prevalent.

Aryn is an {\em unstructured analytics} platform, powered by LLMs, that is designed
to answer these types of queries. We take inspiration
from relational databases, from which we borrow the principles of
declarative query processing. With Aryn, users specify {\em what} they want to ask
in natural language, and the system automatically constructs a plan
(the {\em how}) and executes it to compute the answer from unstructured data.

\begin{figure}[t]
  \centering
  \includegraphics[width=\linewidth]{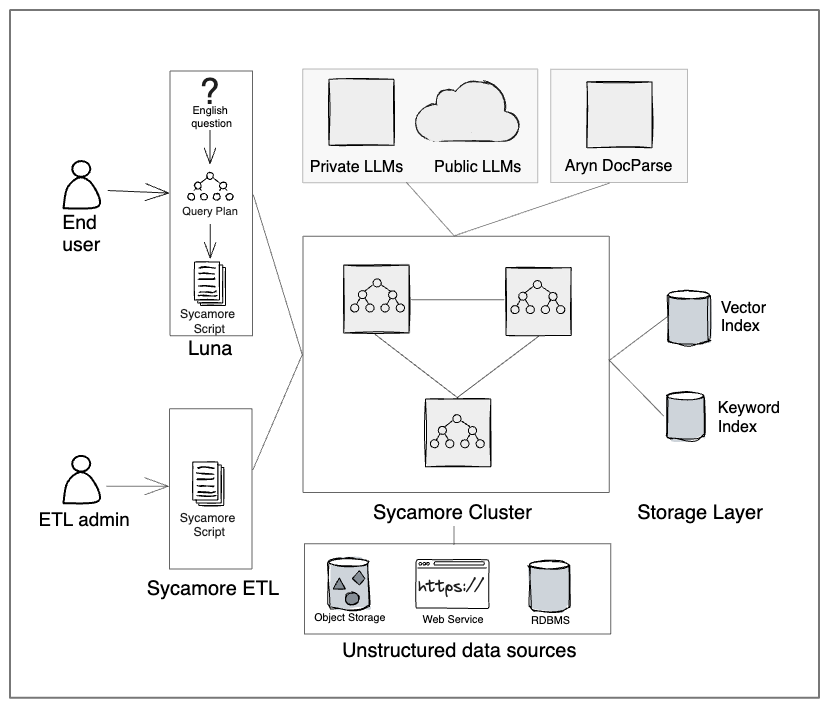}
  \vspace{-10pt}
  \caption{Aryn Architecture}
  \vspace{-10pt}
  \Description{A comparison between the previous and current partitioner.}
  \label{fig:arch}
\end{figure}

Aryn consists of several components (see Figure \ref{fig:arch}). The natural-language query planner,
{\em Luna}, uses LLMs to translate queries to semantic query
plans with a mix of structured and LLM-based semantic
operators. Query plans are compiled to {\em Sycamore}, a document processing
engine used both for ETL and query processing. Sycamore is
built around {\em DocSets}, a reliable abstraction similar to Apache Spark
DataFrames, but for hierarchical documents. Finally, {\em DocParse} uses vision
models to convert complex documents with text, tables, and
images into DocSets for downstream processing.

The main challenge for an analytics system built largely on AI
is to give answers that are accurate and trustworthy. We use LLMs
and vision models for different purposes throughout our stack and
carefully compose them to provide answers to complex questions.
Unfortunately, LLMs are inherently imprecise, making LLM output difficult to verify.

Aryn's database-inspired approach addresses this challenge in multiple ways.
First, our experience working with customers has shown how essential ETL is
for achieving good quality for both RAG and analytics use cases. By performing
high-quality parsing and metadata extraction, we can provide the LLM with the
context necessary to reduce the likelihood of hallucinations.
Second, by dynamically breaking down complex questions into query
plans composed of simple LLM-based operations, we can make the query
plan as a whole more reliable than RAG-based approaches. Third, Aryn exposes the
query plan and data lineage to users for better explainability. A key
component of Aryn is its conversational user interface. Users can
inspect and debug the generated plans, analyze data traces from
execution, and ask follow-up questions to dig in and iterate. This
approach makes it easier for users to navigate their data and helps
build trust in the answers.

In this paper, we describe the motivating use cases for Aryn, tenets driving its design,
and its architecture. We discuss how each component of Aryn works and how they fit together, through an end-to-end use 
case analyzing NTSB\footnote{National Transportation Safety Board (\url{https://www.ntsb.gov/})} incident reports,
which consist of a large collection of unstructured PDF documents containing text, images, figures, and tables.
We also present the user interfaces to inspect, analyze, and debug the plans generated 
by Aryn, and highlight the simplicity of the Sycamore programming framework that makes it easy to 
analyze vast collections of hierarchical unstructured documents. Aryn is fully open source,
Apache v2.0 licensed, and available at \url{https://github.com/aryn-ai/sycamore}.

\section{Use Cases, Challenges, and Tenets}

Enterprises often have document collections with a theme, such as
interview transcripts, earnings reports, insurance
claims, city budgets, or product manuals. Typically,
these documents are a mix of unstructured text with multi-modal data
in tables, graphs, and images. Users often want to
ask complex questions or run analyses that span multiple documents, if
not across a broad subset of the collection.

In working with customers, we see two main classes of use cases:

\vspace{2pt}
\noindent\textbf{Ad-hoc Question Answering:} There's been a recent surge in
AI assistants and chatbots for customer support, typically powered by
technical documentation, ticketing systems like Jira, and internal
messaging boards like Slack. Beyond search-based bots, companies are
also building research and discovery platforms for ad-hoc, chat-driven
analytics. For example, financial and legal firms use such platforms
to aggregate internal research or case summaries and make them
available for analysts to investigate and generate new insights and strategies.

\vspace{2pt}
\noindent\textbf{Report Generation and Business Intelligence (BI):} As LLMs become cheaper and
faster, companies have started building LLM-powered document
pipelines to generate reports. For example, these may be summaries of
hours of user interviews, daily highlights extracted from medical notes
of a patient, or legal claims derived from accident reports.
Going a step further, we see customers extracting structured summary
datasets from document collections to help in critical business
decisions. For example, auto insurance firms want to extract damage
and repair data from claim summaries to understand trends and spot
anomalies as potential fraud.

\vspace{4pt}
\noindent\textbf{Challenges:}
LLMs have inspired many new enterprise use cases because of their
remarkable ability to process unstructured documents. While LLMs hold
promise, they are not enough. LLMs inherently hallucinate, which
is a liability in the use cases described above. In these
settings, users need accurate and explainable answers.

RAG is a popular method to answer questions from documents, but is
fundamentally limited. RAG uses semantic search to retrieve relevant
chunks of documents that are then supplied to an LLM as context to answer
a question. While the RAG approach somewhat mitigates hallucination,
LLM context windows are limited, and
studies show that LLMs with extremely long contexts cannot ``attend''
to everything in the prompt~\cite{lost-in-middle-2020}. RAG
works for simple factual questions where an answer is
contained in a small number of relevant chunks of text, but fails when
the answer involves synthesizing information across a large document
collection.

Another approach is to extract metadata from documents through an ETL
process (perhaps using LLMs) and load it into a database. While this
addresses scale concerns, this does not handle analyses that require
semantic operations at query time.

As an example, consider the question, ``What are the top three most
common parts with substantial damage in accidents involving single
engine aircraft in 2023?''. In NTSB aviation incident reports (about 170K~PDFs
reporting on incidents since 1962), the parts damage details are in free form
text descriptions of the incidents. RAG fails for this because the relevant reports don't
fit into the LLM context.
Moreover, a pure database approach is unhelpful if the fields to be queried have not
been extracted during the ETL phase. In contrast, Aryn generates a plan that quickly
narrows to the relevant incidents in 2023 with a metadata search, and
extracts the parts data at query time using LLM-based semantic
operators.

\vspace{4pt}
\noindent\textbf{Tenets:}
While our approach mitigates concerns of scale, hallucinations, and
reliability, it does not completely eliminate them. We argue that all
systems and approaches built on AI inherently cannot. To address these
challenges, we adopted the following tenets in the design of Aryn.

\begin{itemize}
  \setlength\itemsep{0.75em}
\item{\em Use AI for solutions hard for humans to come by, but easy for humans to verify.}
  The most successful applications of AI have been where AI is used to
  generate solutions and those solutions are verified independently. For
  example, GitHub Copilot generates code, and developers verify its
  correctness in their natural review and testing workflow. Similarly, Aryn uses
  LLMs to generate an initial query plan from natural language, but a
  human is able to inspect and modify the plan if needed.

\item {\em Ensure explainability of results.} Answers to analytics
  questions are hard to verify without manually repeating the
  work. We should make it easy for the user to understand the operation
  of the system and to audit the correctness
  of any result. For example, Aryn provides a detailed trace of how
  the answer was computed, including the provenance of intermediate
  results. In addition, users can ask follow-up questions to navigate the
  results and build trust.

\item {\em Compose narrow AI models and focused AI tasks into a larger whole.}
  Instead of attempting to build the one true model in the vein of
  AGI, we have found it more practical to get reliability and better quality
  if we take a systems approach. In DocParse, we compose a
  variety of vision models for different tasks: segmentation, table
  extraction, and OCR. Similarly, instead of a single LLM invocation
  per query, we break down queries into narrower, more focused
  operators, potentially executed by different LLMs. This improves the
  reliability of each task, and thereby reliability of the whole system.

\end{itemize}

\begin{figure}[t]
  \centering
  \includegraphics[width=\linewidth]{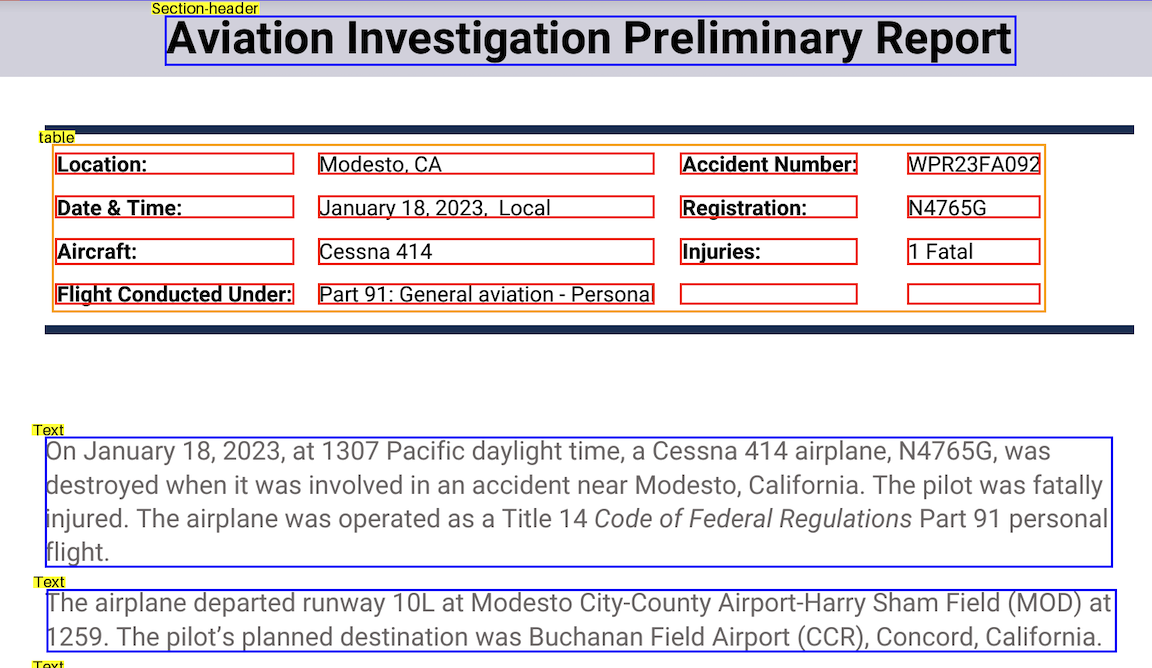}
  \caption{Output of Aryn DocParse (including table and cell identification) on a typical PDF NTSB accident report.}
 \label{fig:ntsb-partition}
\end{figure}

\section{Architecture}

Figure \ref{fig:arch} shows the high level architecture of Aryn. The first
step in preparing unstructured data for analytics is to parse and label raw
documents for further processing. The Aryn \emph{DocParse} service uses modern
vision models to decompose and extract structure from raw documents and
transforms them into DocSets. We developed our own model based on the
deformable DETR architecture~\cite{zhu2020deformable} and trained on
DocLayNet~\cite{doclaynet}.

At the core of Aryn is {\em Sycamore}, a document processing engine that is
built on {\em DocSets}. DocSets are reliable distributed collections, similar to
Spark~\cite{zaharia2012spark} DataFrames, but the elements are hierarchical documents
represented with {\em semantic trees} and additional metadata. Sycamore
includes transformations on these documents for both ETL purposes,
e.g., flatten and embed, as well for analytics, e.g., filter, summarize,
and extract. We use LLMs to power many of these transformations, with
lineage to help track and debug issues when they arise. Sycamore can read data
from a data lake where unstructured data is kept, and can index
processed data in a variety of databases, including keyword and vector stores, 
for use during query processing.

Our query service, {\em Luna}, includes the planner that translates natural
language questions into semantic query plans, which are compiled to Sycamore scripts
for execution. We use LLMs for generating query plans
that users can inspect and validate. This provides explainability for
answers and also allows for debugging and quick iteration. We also use
LLMs for implementing semantic query operators like filtering,
summarization, comparison, and information extraction.

The following sections describe each component in more detail.

\section{DocParse}
\label{sec:partitioner}

One of the lessons we learned early on while building the Aryn system is that
we have to treat data preparation as a key part of any unstructured analytics
system rather than an add on. Parsing complex documents is difficult, and it
is not reasonable to expect users to be able to convert their data into a
text-based format. To address this need, we built the DocParse service to
parse documents and extract information like text, tables, and images.
DocParse exposes a simple REST API that takes a document in a common format
(PDF, DOCX, PPT, etc) and returns a collection of labeled chunks that
correspond to entities in the source document. For example,
Figure~\ref{fig:ntsb-partition} shows a visual representation of how DocParse
parses an NTSB document. It identifies headers, text, and tables, and further
breaks down the structure of the table down to individual cells.

\begin{figure}[t]
  \centering
  \includegraphics[width=\linewidth]{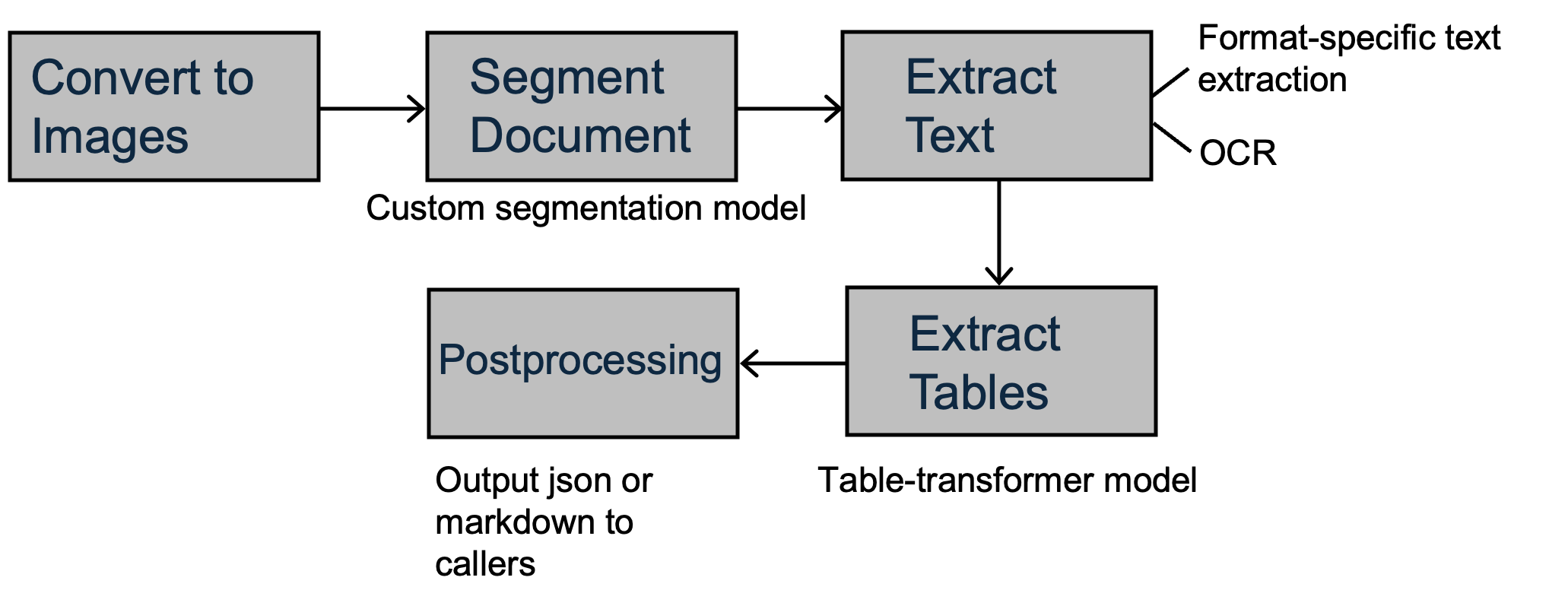}
  \caption{The DocParse Pipeline.}
 \label{fig:table_pipeline}
\end{figure}

DocParse is a compound system composed of multiple stages, as illustrated in
Figure~\ref{fig:table_pipeline}. We first split each document into pages and
convert each page to an image so that we can leverage vision models in later
stages of the pipeline. This approach, which is commonly used in document
processing, allows us to support a variety of formats in a consistent way and
take advantage of semantic information in the rendered document that may be
difficult to extract from the underlying file format such as the relative size
or position of objects on the page.

The next step in the pipeline is segmentation, which uses an object detection
model to identify bounding boxes and label them as one of 11 categories,
including titles, images, paragraphs, and tables. As we were developing
DocParse, we found that many of the existing open source object detection
models performed poorly on document segmentation, so we trained our own. We
used the Deformable DEtection TRansformer (DETR)
architecture~\cite{zhu2020deformable} and trained it on
DocLayNet~\cite{doclaynet}, an open source, human-annotated document layout
segmentation dataset. We have made this model available for use with a
permissive Apache v2 license on Hugging Face~\cite{deformable-model-hf} and
have continued to update the version used by DocParse by collecting and
labeling customer documents. We evaluate the performance of this model in
Section~\ref{sec:partition_eval}.

The segmentation model outputs labeled bounding boxes, but it doesn't have any
information about the text in the document. The next stage in the pipeline is
to extract this text. Depending on the document format, this is done by
reading text directly from the underlying file format with a tool like
PDF~Miner\footnote{\url{https://pypi.org/project/pdfminer/}} or with an OCR
tool like EasyOCR\footnote{\url{https://github.com/JaidedAI/EasyOCR}} or
PaddleOCR\footnote{\url{https://github.com/PaddlePaddle/PaddleOCR}}. Once we
have have the text and the labeled bounding boxes, we can perform additional
type-specific processing. For instance, for tables, we use a Table
Transformer-based model~\cite{Smock_2022_CVPR} to identify the
individual cells, while for images we can use a multi-modal LLM to compute a
textual summary.

As part of post-processing, DocParse combines the output from each page into a
final result, either in JSON or a higher-level format like Markdown. Users can
leverage Sycamore to import and manipulate the JSON directly and perform more
complex data processing transformations.

\subsection{Evaluation}
\label{sec:partition_eval}

\begin{table}
\centering
\small
\begin{tabular}{|l|r|r|}
    \hline
    \textbf{Service} & \textbf{mAP} & \textbf{mAR}\\
    \hline
    DocParse & \textbf{0.640} & \textbf{0.747} \\
    \hline
    Amazon Textract~\cite{amazon_textract} & 0.423 & 0.507 \\
    \hline
    Unstructured (REST API with YoloX)~\cite{unstructured_api} & 0.347 & 0.505 \\
    \hline
    Azure AI Document Intelligence~\cite{azure_doc_intelligence} & 0.266 & 0.475 \\
    \hline
\end{tabular}
\caption{Segmentation performance on the DocLayNet competition benchmark~\cite{partition_blog}}.
\label{tab:partition_results}
\end{table}

In order to evaluate the performance of our segmentation model, we used the
DocLayNet competition benchmark~\cite{doclaynet-comp}. This benchmark was
developed by the authors of the DocLayNet dataset, it includes documents drawn
from a variety of domains, including those not directly represented in the
training dataset. The evaluation is done using the standard COCO
framework~\cite{coco}, which measures mean average precision (mAP) and mean
average recall (mAR) across the 11 DocLayNet object classes.
Table~\ref{tab:partition_results} shows the a comparison of DocParse
against several other document processing services. In order to make the
comparison as fair as possible, we standardized the set of labels across all
four services, and removed results containing labels that were not present in
one or more of the services. More information on our methodology can be found
in the corresponding blog post~\cite{partition_blog}. Our results show that
DocParse is between 1.5 and 2.4 times more accurate than competing
services in terms of mAP, and between 1.5 and 1.6 times more accurate in terms
of mAR. These results validate our approach and suggest that DocParse can
serve as the first step towards ingesting documents into an unstructured
analytics system.

\section{Sycamore}
\label{sec:sycamore}

Sycamore is the open-source document processing engine at the center of the Aryn system~\cite{sycamore}.
We built Sycamore to support both data preparation and analytics over complex document sets.
One of the primary motivations for Sycamore was the observation that the
line between ETL and analytics gets blurred when dealing with unstructured data. In particular,
we need the flexibility to run certain document processing operations either at ETL time
or at query time. For example, the cost of an expensive LLM-based processing step can be
amortized over many queries by running it once during ETL, but because the space of
potential queries is very large, not all operations can be performed in advance.

To accommodate these challenges, we built Sycamore as a dataflow system inspired by Apache
Spark~\cite{zaharia2012spark}, with extensions to integrate with LLMs and support
unstructured documents.

\subsection{Data Model}


Documents in Sycamore are hierarchical and multi-modal.
A long document may have chapters that are broken into sections, which in turn contain individual
chunks of text, or entities like tables and images. The latter data types are particularly important
for many analytics queries and need special treatment. More precisely, a document in Sycamore is a tree,
where each node contains some content, which may be text or binary, an ordered list of child nodes, and
a set of JSON-like key-value properties. We refer to leaf-level nodes in the tree as \emph{elements}.
Each element corresponds to a concrete chunk of the document and is identified as one of 11 types,
such as a text, image, or table. Each element may have special reserved properties based on its type. For example, a TableElement has properties containing rows and columns, while an ImageElement has information about format
and resolution. DocSets are flexible enough to represent documents at different stages of processing.
For example, when first reading a PDF, it may be represented as a single-node document with the
raw PDF binary as the content. After parsing, each section is an internal node and tables and
text are identified as leaf-level elements.

\subsection{Programming Model and Operators}
\begin{table}[t]
\small
\subfloat[Structured operators in Sycamore]{%
\begin{tabularx}{\linewidth}{ |l|X| }
  \hline
  \texttt{queryDatabase} & Scans documents from an index based on keyword search over the element content or filters over the properties. \\
  \hline
  \texttt{map}, \texttt{filter}, \texttt{flatMap} & Transforms documents using standard functional operators. \\
  \hline
  \texttt{partition} & Parses a document using DocParse. \\
  \hline
  \texttt{explode} & Unnests each element and makes it a top-level document. \\
  \hline
  \texttt{reduceByKey} & Standard reduce operation that can be used for a variety of grouping and aggregations on properties on the documents. \\
  \hline
  \texttt{write} & Writes a DocSet to a database.\\
  \hline
  
\end{tabularx}%
}\\

\subfloat[Semantic operators in Sycamore]{%
\begin{tabularx}{\linewidth}{ |l|X| }
  \hline
  \texttt{queryVectorDatabase} & Performs semantic search over a collection of indexed documents, returning a DocSet with the matches.\\
  \hline
  \texttt{llmFilter} & Uses an LLM prompt to drop or retain documents in a DocSet.\\
  \hline
  \texttt{llmExtract} & Extracts one or more fields from each document using an LLM, saving the results as document properties. \\
  \hline
  \texttt{llmReduceByKey} & Similar to \texttt{reduceByKey}, but uses an LLM to combine multiple documents.\\
  \hline
  \texttt{embed} & Computes embeddings for each document. \\
  \hline
\end{tabularx}%
}
\caption{Example Sycamore Operators}
\label{tab:sycamore_operators}
\end{table}


Programmers interact with Sycamore in Python using a Spark-like model of
functional transformations on DocSets. Table~\ref{tab:sycamore_operators} shows
several of Sycamore's operators. We classify these operators as either
\emph{structured} or \emph{semantic}. Structured operators correspond to
standard dataflow-style operations. These include functional operators like
\texttt{map} and \texttt{filter} that take in arbitrary Python functions, as well
as transformations like \texttt{partition} and \texttt{explode} that modify the structure of
documents by creating or unnesting elements, respectively. The
\texttt{reduceByKey} operation makes it possible to support map-reduce style
operations and implement aggregation by document properties. These transforms
accommodate the fact that some documents may be missing certain fields. Sycamore does not yet
support full joins.

Semantic operators leverage LLMs to perform transformations based on the
content or meaning of documents. These operators are often driven by natural
language prompts and are typically used to enrich document metadata. Many of
the semantic operators, like \texttt{llmFilter}, can be implemented in terms
of the structured operators. We still prefer to separate them out because they
can behave very differently in practice, as LLMs are inherently non-deterministic and users
often want to manually inspect the results of semantic operations. Sycamore
supports a variety of LLMs, including those from OpenAI and Anthropic, and
open source models like Llama.

The code\footnote{We have elided a few configuration parameters to enhance readability.}
in Figure~\ref{fig:syc_code} is an example of processing NTSB incident report documents using Sycamore.
The code partitions documents using DocParse, described in Section~\ref{sec:partitioner}.
It then executes the \texttt{llmExtract} transform,
which takes a JSON schema and attempts to extract those fields from each document using an LLM.
As shown in in Figure~\ref{fig:entity_extract}, this approach correctly extracts the state abbreviation
and other fields from the document. Next, we use \texttt{explode} to break each document into a
collection of document chunks, and then we generate an embedding vector for each chunk.
At this point the DocSet is ready to be loaded into a database like OpenSearch 
for later querying (using \texttt{write}).

Finally, \texttt{queryDatabase} and \texttt{queryVectorDatabase} support reading a previously loaded
DocSets from a data store. The \texttt{queryDatabase} operator is analogous to a standard
database scan operator, and supports filters on the metadata as well as keyword search (depending on the
capabilities of the data store). The \texttt{queryVectorDatabase} operator, in addition to those, also 
supports {\em semantic} search (i.e., {\em vector similarity} search) over the chunks. While indexing is
done on chunks, Sycamore reassembles these chunks into documents before passing them to downstream operators. 


\begin{figure}
\small
\begin{verbatim}
schema = {
  "us_state": "string",
  "probable_cause": "string",
  "weather_related": "bool"
}

ds = context.read.binary("/path/to/ntsb_data")
      .partition(DocParse())
      .llmExtract(
        OpenAIPropertyExtractor("gpt-4o", schema=schema))
      .explode()
      .embed(OpenAIEmbedder("text-embedding-3-small"))
\end{verbatim}
\caption{Sample Sycamore script.}
\label{fig:syc_code}
\end{figure}

\begin{figure}
  \small
\begin{verbatim}
{
  "us_state_abbrev":"AK",
  "probable_cause": "The pilot's failure to remove
    all water from the fuel tank, which resulted in fuel
    contamination and a subsequent partial loss of engine power.",
  "weather_related': True
}
\end{verbatim}
\caption{Output of the \texttt{llmExtract} transform.}
\label{fig:entity_extract}
\end{figure}

\subsection{Execution}

Sycamore adopts a Spark-like execution model where operations are pipelined and
executed only when materialization is required. To assist with debugging and
avoid redundant execution, Sycamore also supports a flexible \emph{materialize}
operation that can save the output of intermediate transformations to memory,
disk, or cloud storage. Sycamore is built on top of the Ray compute framework~\cite{moritz2018ray},
which provides primitives for running distributed Python-based dataflow workloads. We chose
Ray because it is based on Python, which has become the language of choice for machine
learning applications, and because it is well-integrated with existing ML libraries.

\section{Luna}
\label{sec:luna}

A hallmark of relational databases is declarative query processing, which
hides the low-level details of how queries are executed and makes it easier
for application developers to adapt to changing workloads and scale. LLMs make
it possible to leverage declarative query processing for {\em natural
language} queries over complex, unstructured data. We call this {\em
LLM-powered unstructured analytics}, or {\em Luna} for short.

More specifically, Luna converts a natural language query into a query plan
that runs over DocSets and returns either raw tabular results or natural
language answers. Query plans are executed using Sycamore's DocSet
operators. To aid explainability, Luna exposes the logical query plan, data lineage, and
execution history, and allows users to modify any part of the plan to better
align with their intention. The remainder of this section describes the
system in detail.

\subsection{Luna Architecture}

Luna consists of a number of pieces that work together to provide an
end-to-end natural-language query processing system over complex, unstructured data.

\vspace{5pt}
 \noindent
 \textbf{Data Inputs and Schema.} Luna shares the Sycamore data model and
 executes queries against one or more DocSets that have been indexed in a
 database. During query planning, we provide the planner with
 the schema of each DocSet, which consists of the properties contained in the
 documents, along with their data types and sample values, along with a special
 ``text-representation'' field representing the entire contents of each
 Document. The schema of DocSets can evolve over time, based on new semantic relationships 
 discovered in the data, potentially driven by the query workload.

 While Sycamore represents documents hierarchically with
 elements corresponding to document chunks, we found it more effective to
 hide this from the planner and always provide a schema for complete
 documents. The Sycamore engine
 handles splitting documents into chunks that fit in to the context window of
 the LLM used for embedding and reconstructing the full document during queries.

 In our implementation, we primarily use OpenSearch for storing and querying
 DocSets, though other data management systems can be used as long as they
 support both ``keyword'' and ``semantic search'' (i.e., vector similarity
 queries) and basic filtering by properties.

\vspace{5pt}
\noindent
\textbf{Logical Query Operators.} 
Luna uses an LLM for interpreting a natural language user query. We initially provided the LLM the complete
list of physical operators as part of the prompt.
However, in our experiments with several real-world datasets and query workloads, we found that this
approach does not work well for complex and exploratory analysis queries like: ``Analyze maintenance-related
incidents by grouping those by aircraft type and maintenance interval to find patterns of recurring issues.''
In particular, we found it difficult to get the LLM to use grouping operations like \texttt{reduceByKey}
effectively
and the plans generated would often run into context window size limitations.

Instead, we decided to differentiate between logical and physical operators with respect to query planning
and execution. Luna provides a simpler set of high-level logical operators to the LLM for query planning
purposes, and rewrites the resulting logical plan into physical operators before execution.
This also makes it easier for the user to understand the plan and debug the execution.

Many simple logical operators map one-to-one to physical Sycamore operators, including single-pass
per-document operations like \texttt{map}, \texttt{filter}, and \texttt{llmExtract}, but for operations
that span multiple documents, we have found it often works better to have more specific operators rather
than low-level primitives. For example, the following logical operators are exposed to the Luna planner:

\begin{itemize}
\item \texttt{groupByAggregate}: Performs a database style group-by and aggregation. 

\item \texttt{llmCluster}: Clusters documents using $k$-means based on semantic similarity of one or more fields. 

\item \texttt{llmGenerate}: Summarizes one or more documents based on a prompt. This is analogous to the ``G'' in ``RAG'' and is often used at the end of a plan. 
\end{itemize}

Each of these operators can be implemented in terms of the existing Sycamore physical operators.
For instance, \texttt{groupByAggregate} and \texttt{llmCluster} can be implemented with a combination
of \texttt{map} and \texttt{reduce} operations, but we see better results from the planner when we keep
them as separate operators.

\begin{figure}
    \includegraphics[width=0.40\textwidth]{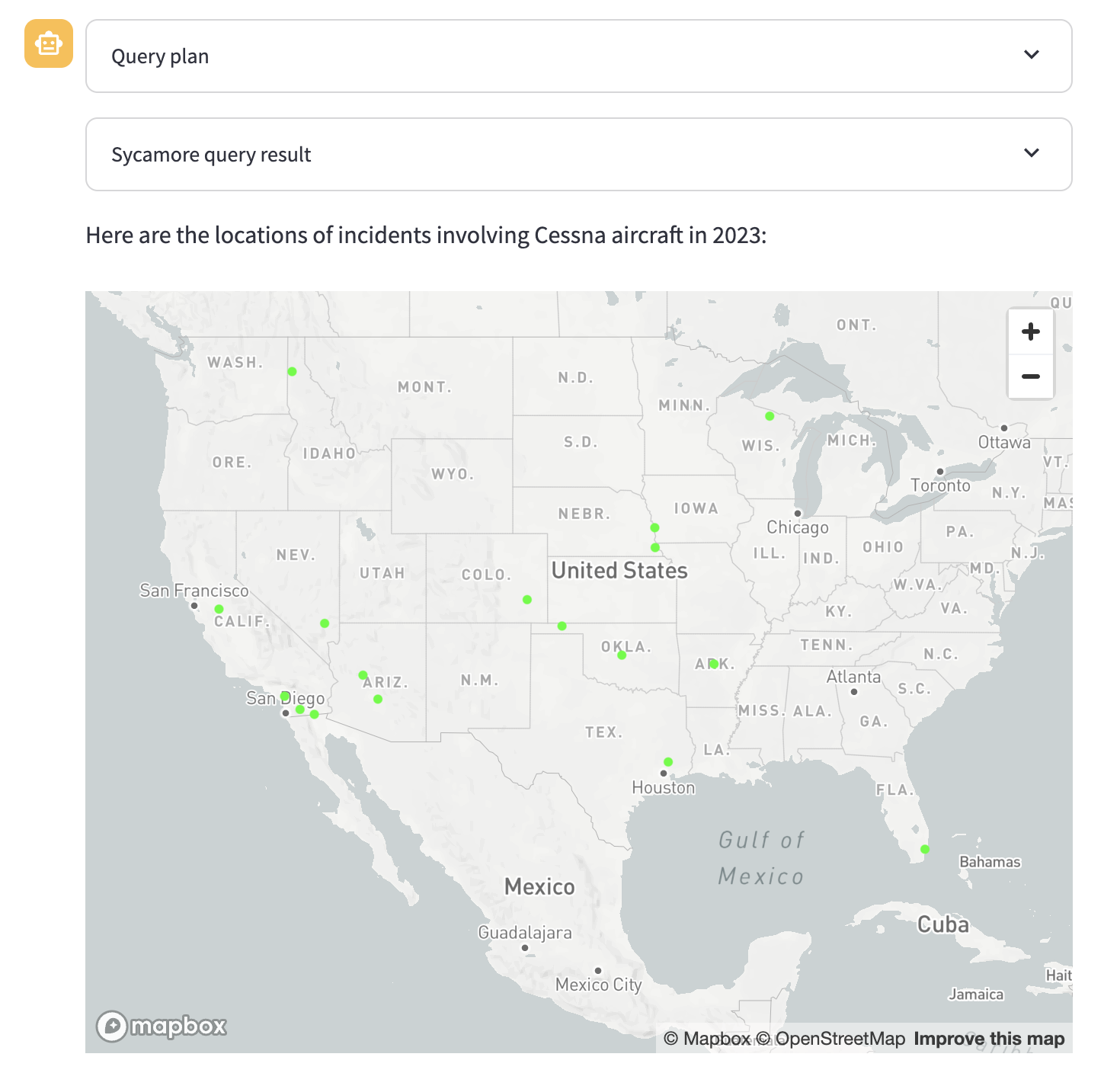}
    \includegraphics[width=0.40\textwidth]{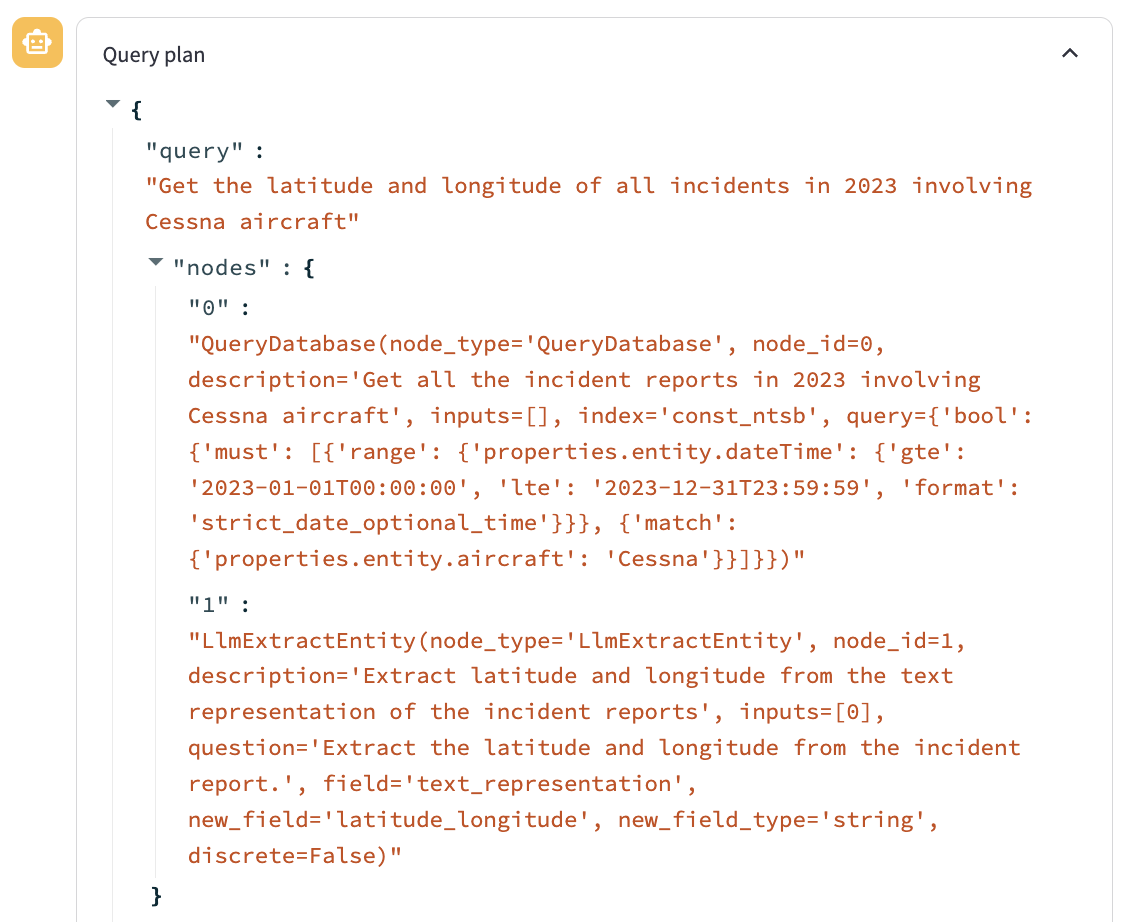}
    \includegraphics[width=0.40\textwidth]{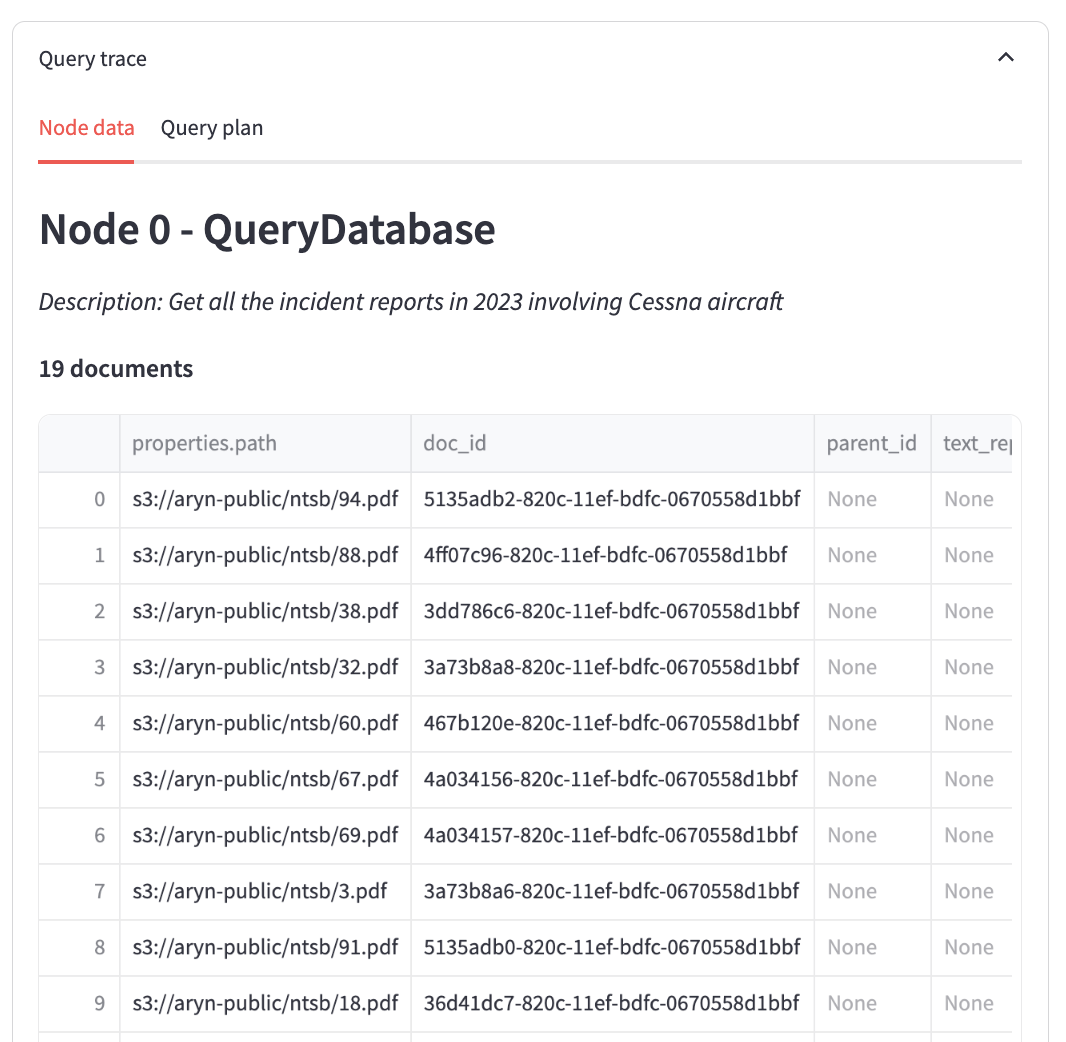}
    \caption{The Luna user interface shows the query result visually, allows the use to inspect the generated query plan, and lets them drill down to individual documents if needed}
    \label{fig:luna:json}
\end{figure}

\vspace{5pt}
\noindent
\textbf{Query Planning.} Luna uses an LLM to interpret a natural language query
and decompose it to a DAG of logical query operators. After significant experimentation, we found that 
including the following information in the prompt helps provide the LLM with the right context:
 \begin{itemize}
    \item The schema for the input DocSet. For each schema field, we include a short description as well
    as a few example values drawn from the underlying data.
    \item A list of available logical operators and their syntax.
    \item A list of example queries and their associated query plans.
 \end{itemize}
We instruct the LLM to generate the plan in JSON format, which we validate against a schema to ensure
that it conforms to the expected syntax. In addition to confirming that the query plan is syntactically
correct, we also check that it is semantically valid. For example, if a QueryDatabase operation performs
field-based filtering, we check that the fields used in the filter are valid for the given DocSet.

\vspace{5pt}
\noindent
\textbf{Plan Rewriting and Optimization.}
Despite significant prompt engineering, the LLM may still produce a suboptimal or, in some cases, 
an incorrect or infeasible query plan. 
We use a combination of plan rewriting and rule-based optimization to address these issues. For example, if the
plan has multiple \texttt{llmExtract} operators in sequence, these can combined into a single
operator.

\vspace{5pt}
\noindent
\textbf{Execution.} After plan rewriting and optimization, the query plan is compiled into Sycamore 
code in Python. Execution on large datasets benefits from 
distributed processing, and using Sycamore's distributed execution
mode allows us to scale out workloads with minimal overhead. The compiled query execution
code in Sycamore is easy for a technically savvy user to understand and modify (in the UI itself).

\vspace{5pt}
\noindent
\textbf{Traceability and debugging.} The ambiguous nature of some 
queries can result in Luna misinterpreting the user's intention. 
It is critical to allow the user to inspect the query execution trace
and provide feedback to correct itself. With a combination
of logging and exposing APIs that allow the user to modify any stage of 
query execution, users have full control over how their query is answered.

\subsection{User Interface and Verifiability}
Luna's user interface, shown in Figure~\ref{fig:luna:json} is designed to make
it easy for users to verify the results from the system. 
Luna achieves this by: (a) exposing the query plan, (b) allowing the user to inspect
intermediate results, and (c) allowing the user to ask follow-up questions to guide the system.

Luna exposes the plan generated from a user query as a simple JSON object. This allows a user to
understand the exact operations that were performed to answer a query, how the dataset was
transformed during each operation, and modify any part of the plan to better align with their intention.
Given the query ``Get the latitude and longitude of all incidents in 2023 involving Cessna aircraft,''
we can see the resulting plan as a \texttt{queryDatabase} operation followed by a \texttt{llmExtract} operation.
The Luna UI also shows the user the Sycamore code that was generated for the query, which they can edit and re-run. 

While inspecting the query plan is often enough to convince oneself that the data generated by the
query is likely to be correct, further validation is possible by inspecting the data flowing out of
each of the operators. The Luna UI allows the user to explore the raw data at each stage of the
query plan, drilling down to individual records and linking back to the original source documents.

Finally, we find that supporting an iterative, exploratory mode of interaction with the system
is essential. Users can test hypotheses and explore different aspects of the data by asking
follow-up questions, such as ``what about incidents without substantial damage'' or 
``show only results in California.''
The conversational history with the system allows a user to refer to previous queries or results
implicitly, making this interaction much more natural, much like asking questions of a human
analyst.

\section{Evaluation}
\label{sec:eval}

We present a preliminary evaluation of Luna's ability to answer complex
analytical questions over a dataset of incident reports from the National
Transportation Safety Board, which is the US-based agency responsible for
investigating civil transportation accidents. Our test dataset consists of~100~PDF
reports pulled from the NTSB CAROL database\footnote{\url{https://carol.ntsb.gov/}}
covering aviation incidents between June and September~2024. Each file is
between 4~and~7~pages of text, with sections covering a summary of the incident,
probable cause and findings, factual information, and administrative information.
Incident reports have multiple tables covering aspects such as the pilot's background,
aircraft and operator details, meteorological information, wreckage, and injuries.
Many of the documents contain photographs of the accident site or maps of the
flight trajectory.

\begin{table}
    \small
    \begin{tabularx}{0.5\textwidth}{|l|X|X|}
        \hline
        \textbf{Field} & \textbf{Example value} \\
        \hline
        \texttt{accidentNumber} & CEN23FA095 \\
        \texttt{aircraft} & Piper PA-38-112 \\
        \texttt{aircraftDamage} & Destroyed \\
        \texttt{conditionOfLight} & Dusk \\
        \texttt{conditions} & Visual (VMC) \\
        \texttt{dateAndTime} & June 28, 2024 19:02:00 \\
        \texttt{departureAirport} & Winchester, Virginia (OKV) \\
        \texttt{destinationAirport} & Yelm; Washington \\
        \texttt{flightConductedUnder} & Part 137: Agricultural \\
        \texttt{injuries} & 3 Serious \\
        \texttt{location} & Gilbertsville, Kentucky \\
        \texttt{lowestCeiling} & Broken / 5500 ft AGL \\
        \texttt{lowestCloudCondition} & Scattered / 12000 ft AGL \\
        \texttt{operator} & Anderson Aviation LLC \\
        \texttt{registration} & N220SW \\
        \texttt{temperature} & 15.8C \\
        \texttt{visibility} & 7 miles \\
        \texttt{windDirection} & 190° \\
        \texttt{windSpeed} & 19 knots gusting to 22 knots \\
        \hline
    \end{tabularx}
    \caption{Schema extracted from NTSB incident reports.}
    \label{tab:ntsb-schema}
    \end{table}

We processed the NTSB~reports using a Sycamore pipeline. The pipeline starts by calling
DocParse to parse each document as described in Section~\ref{sec:partitioner},
and then uses the \texttt{llmExtract} transform to extract key data from each
document. We load the resulting schema, shown in Table~\ref{tab:ntsb-schema},
into an OpenSearch index. We also chunk and embed the text content of the incident reports, and
the resulting vectors are also stored in OpenSearch for use with vector search operations.
Throughout this evaluation we used OpenAI's gpt-4o model for our LLM, all-MiniLM-L6-v2 for the embeddings, and OpenSearch 2.17.

\subsection{Benchmark questions}

There does not exist a standard benchmark for document analytics against this type of dataset.
Through manual inspection, we derived a set of 30~questions that represent a broad
range of query types and varying degrees of difficulty to answer. Some examples of the
benchmark questions include:
\begin{itemize}
    \item How many incidents were there by state?
    \item What fraction of incidents that resulted in substantial damage were due to engine problems?
    \item In incidents involving Piper aircraft, what was the most commonly damaged part of the aircraft?
    \item Which incidents occurred in July involving birds?
\end{itemize}

A few of the benchmark questions can be answered more or less directly by querying the
extracted metadata shown in Table~\ref{tab:ntsb-schema}. However, in most cases, the benchmark questions
refer to information not explicitly captured in the schema, such as whether an incident
involved birds or engine problems. For these cases, Luna needs to use a
combination of metadata lookup and LLM-based extraction or filtering based on the documents'
textual content.

Many of our benchmark questions would be difficult, or impossible, for a RAG-based system to
answer, given that the information required to answer the question is spread across multiple
portions of each document, and a vector search would not be expected to return meaningful chunks of
context for downstream analysis by the LLM.

\subsection{Results}

We ran Luna against each of our 30~benchmark questions and compared the result
to ground truth answers determined through manual inspection. As a comparison point, we also
used RAG to answer each question, using a standard RAG approach that first converts the
question into a vector search against the embedded set of text chunks, retrieves the $k$
nearest documents for each question, and provides those chunks as context to the LLM to answer
the original question. For this test we set $k=100$. The results are shown in Table~\ref{tab:eval-results}.

\begin{table}
    \small
    \begin{tabularx}{0.5\textwidth}{|l|X|X|}
        \hline
        & \textbf{Luna} & \textbf{RAG} \\
        \hline
        Correct & 20 (67\%) & 2 (6.7\%) \\ \hline
        Incorrect & 10 (33\%) & 20 (67\%) \\ \hline
        Refusal & 0 (0\%) & 8 (26.7\%) \\ \hline
        {\bf Total} & {\bf 30} & {\bf 30} \\ \hline
    \end{tabularx}
    \caption{Luna vs. RAG evaluation results on NTSB document questions.}
    \label{tab:eval-results}
\end{table}

Luna answers 20 out of the 30 questions correctly, and 10 incorrectly. The incorrect
answers fall into several categories:
\begin{description}
\item[Counting errors (6 cases).] In several cases, there are off-by-one errors due to incidents
being counted twice. For example, for the question ``How many incidents were there, broken down
by number of engines?'', there is a single incident involving two aircraft, each with 1 engine.
These are counted as two separate ``incidents''. Fixing this would require a deduplication
step in the query plan which can be achieved with better few-shot examples for the planner.
\item[Filter errors (3 cases).] The LLMFilter operation is occasionally too generous
in its interpretation of whether a given document should pass the filter test. As an example,
in the question ``How many incidents were due to engine problems?'' the LLM filter operation
screens for ``Does the document indicate engine problems?''. Because portions of most NTSB
reports mention engines in various contexts, the filter tends to pass through documents
where an engine {\em problem\/} was not indicated. Better prompting for the filter conditions
would help here.
\item[Query interpretation (1 case).] For the question ``What was the breakdown of incident types
by aircraft manufacturer?'', the LLM interprets ``aircraft manufacturer'' to mean whether the
aircraft was military, commercial, a helicopter, or some other type, rather than the name of
the manufacturer (which is indeed present in the dataset). This would be fixable with some
additional few-shotting, but points more broadly to the challenge of teaching the LLM about
the semantic interpretation of the schema.
\end{description}

As we expected to see, RAG does poorly on most of these questions. The two cases
in which RAG gets the correct answer are ``How many incidents were there in Hawaii?''
(for which the correct answer is zero), and ``Which incidents occurred in July involving
birds?'' (two incidents). Both of these are answerable using the RAG approach when
the number of records retrieved from the vector search is small enough to fix in the LLM's
context window. RAG does not yield the correct answer in any case where the number of matching
incidents exceeds a modest threshold, such as ``How many incidents involved substantial damage?''
(correct answer: 94, RAG answer: 10). 

A substantial number of RAG queries resulted in a refusal of the LLM to answer the question
at all. For example, on the question ``How many incidents were due to engine problems?'', the
LLM responds with ``The NTSB does not assign fault or blame for accidents or incidents, including
those related to engine problems.'' This is caused by {\em context poisoning\/} during the
RAG process. Each of the NTSB reports contains a boilerplate disclaimer that states,
\begin{quote}
``The NTSB does not assign fault or blame for an accident or incident; rather, as specified
by NTSB regulation, 'accident/incident investigations are fact-finding proceedings with no
formal issues and no adverse parties ... and are not conducted for the purpose of determining
the rights or liabilities of any person' (Title 49 Code of Federal Regulations section 831.4).''
\end{quote}
Whenever these text chunks are included in the vector search results fed as context to the
LLM, the final response is effectively poisoned by the disclaimer. While this could be addressed
though a range of prompting and santization techniques, we chose to highlight this as an
interesting failure mode of the conventional RAG approach.

\section{Related Work}
Machine learning has revolutionized many aspects of data management over the last decade.
First, there is a long line of work on natural language to SQL~\cite{kim2020natural,pourreza2024din, chase-sql2024}.
While the early work focused on building specialized models for this purpose, LLM-based approaches have proven superior in recent years\footnote{See the leaderboard at \url{https://yale-lily.github.io/spider}.}.
Several recent works have focused on generating queries that incorporate LLM calls~\cite{liu2024suql,liu2024declarative,urban2023caesura}.
Our Luna framework is differentiated by a broader set of LLM-based 
operations, a focus on hierarchical documents, and our emphasis on interactive interfaces.

There is also much work on using LLMs for specific ETL tasks such as entity resolution, 
information extraction, named entity recognition, and data cleaning~\cite{li2020deep,narayan2022can,suhara2022annotating}.
In addition, there's also work in detecting and extracting tables using modern transformer models~\cite{Smock_2022_CVPR, unitable-2024}, OCR~\cite{paddleocr-2022, easyocr}, and segmentation and labeling~\cite{doclaynet,doclaynet-comp}. To date, ours is the only work that combines the best of these
into a unified cloud service and is deeply integrated with a declarative document processing framework for ETL like Sycamore.

DocParse is based on a long line of work in document segmentation. Current approaches commonly use
object detection models such as DETR~\cite{detr}. DocParse follows this approach and
leverages Deformable DETR~\cite{zhu2020deformable}. An
alternate line of work has led to multi-modal models such as Donut~\cite{kim2021donut} and
LayoutLMV3~\cite{huang2022layoutlmv3} that seek to directly solve document understanding tasks like visual question
answering (VQA) without the need for explicit segmentation. Sycamore can eventually incorporate these models, but we
continue to find segmentation valuable as we can index the segments to reduce work at query time.

There is less work on building end-to-end systems that encompass the entire spectrum of tasks from document parsing to ETL to querying for unstructured document analytics. Nonetheless, 
several similar efforts have 
started over the last year
including ZenDB~\cite{zendb}, LOTUS~\cite{patel2024lotus}, EVAPORATE~\cite{arora2023language},
CHORUS~\cite{kayali2024chorus}, and Palimpzest~\cite{liu2024declarative}.
TAG~\cite{tag-2024} is similar in spirit to Luna, but translates to SQL and does not include LLM-based operators post database query.
Most recently, DocETL~\cite{docetl-2024} proposes to use agent-based rewrites to automatically optimize document processing pipelines for improved accuracy.
In contrast to these works, while we have incorporated similar pipelining and rewriting mechanisms to start, we do not believe it is possible to fully automate and optimize the entire pipeline in practice.
As a result, we have designed Aryn to facilitate a human-in-the-loop paradigm.

\section{Conclusions and Future Work}
We are building Aryn to make unstructured data as easy to query as structured data by leveraging the immense potential of LLMs to process multi-modal datasets.
We take a database-inspired approach of decomposing analytics queries into semantic query plans which not only improves answer accuracy but also provides explainability and an avenue for intervention and iteration. At the same time, given the limitations of current models, we are building Aryn to be a human-in-the-loop system; as the LLMs improve, the need for human interventions will diminish, but it is unlikely to completely vanish.
Our experience across a variety of application domains supports that our overall design as well as Aryn's individual components are promising.
Nonetheless, many challenges still remain.
We need to continue to improve accuracy and make it easier to adapt Aryn to new use cases.
We need ways to correct and evolve the system and automatically learn from  users as they exercise the system.
We need to extend Aryn to support joins and allow queries to incorporate external sources like data warehouses. 
Finally, we've just started the journey on improving performance, cost, and scale.




\begin{acks}
We thank Amol Deshpande for his insights, detailed advice, and contributions from the start and throughout our journey at Aryn.
We also thank our newest members who relentlessly make the platform better: Akarsh Gupta, Soeb Hussain, Dhruv Kaliraman, Soham Kasar, Abijit Puhare, Karan Sampath, Aanya Pratapneni, and Ritam Saha.
We are indebted to our customers whose partnership makes our contribution unique and differentiated.
Finally, we thank our reviewers for their suggestions.
\end{acks}

\bibliographystyle{ACM-Reference-Format}
\bibliography{main.bib}

\end{document}